\documentclass[12pt,a4paper,final]{iopart}

\usepackage{iopams}  
\usepackage{graphicx}
\usepackage[breaklinks=true,colorlinks=true,linkcolor=blue,urlcolor=blue,citecolor=blue]{hyperref}
\usepackage{amssymb}
\usepackage{graphicx}	
\usepackage{dcolumn}	
\usepackage{bm}				
\usepackage[]{units}	
\usepackage{color}
\usepackage{textcomp}
\usepackage{eqnarray}
\usepackage{booktabs}
\usepackage{multirow}
\usepackage{dcolumn}

\begin{document}

\title[]{Thermometry of Guided Molecular Beams from a Cryogenic Buffer-Gas Cell}

\author{X.~Wu, T.~Gantner, M.~Zeppenfeld, S.~Chervenkov, and G.~Rempe} 

\address{Max-Planck-Institut f\"ur Quantenoptik, Hans-Kopfermann-Str. 1, D-85748 Garching, Germany}

\begin{abstract}

We present a comprehensive characterization of cold molecular beams from a cryogenic buffer-gas cell, providing an insight into the physics of buffer-gas cooling. Cold molecular beams are extracted from a cryogenic cell by electrostatic guiding, which is also used to measure their velocity distribution. Molecules' rotational-state distribution is probed via radio-frequency resonant depletion spectroscopy. With the help of complete trajectory simulations, yielding the guiding efficiency for all of the thermally populated states, we are able to determine both the rotational and the translational temperature of the molecules at the output of the buffer-gas cell. This thermometry method is demonstrated for various regimes of buffer-gas cooling and beam formation as well as for molecular species of different sizes, CH$_3$F and CF$_3$CCH. Comparison between the rotational and translational temperatures provides evidence of faster rotational thermalization for the CH$_3$F-He system in the limit of low He density. In addition, the relaxation rates for different rotational states appear to be different.

\end{abstract}

\vspace{2pc}

\maketitle

\section{Introduction}
Cold molecules are intriguing quantum objects with a complex energy-level structure, yet amenable to excellent control over their motional and internal degrees of freedom~\cite{Carr2009,Lemeshko2013}. This renders them particularly suitable for studying low-energy collision dynamics~\cite{Gilijamse2006,Parazzoli2011,Henson2012,Kirste2012,Strebel2012,Hall2012,Tariq2013,Hauser2015} and enabling controlled chemistry~\cite{Tscherbul2006a,Ni2010}. Cold molecules are also beneficial for precision spectroscopy, serving as an important tool for exploring fundamental physics~\cite{Hudson2011,ACME2014,Dietiker2015}. For various applications of cold polar molecules, achieving high purity of and control over their internal states~\cite{Viteau2008,Manai2012,Gloeckner2015}, in addition to having the ability to manipulate their motional behaviour~\cite{Gupta1999,Bethlem1999,Fulton2004,Narevicius2008b,Hogan2009,Sommer2010,Shuman2010,Zeppenfeld2012,Chervenkov2014}, is of paramount importance. In pursuit of this goal, cryogenic buffer-gas cooling has proven to be a very general and powerful method to produce internally and translationally cold molecules~\cite{Weinstein1998,Maxwell2005,vanBuuren2009}.\\
Here we demonstrate the full characterization of the properties of buffer-gas-cooled molecular beams, and, on this basis, elucidate various aspects of buffer-gas cooling. Our characterization relies on the combination of four components. First, we extract cold molecules from a buffer-gas cell by electrostatic guiding in a quadrupole guide~\cite{vanBuuren2009}. Second, time-of-flight measurements are used to derive the longitudinal velocity distribution of the molecules at the end of the guide~\cite{Sommer2009}. Third, resonant radio-frequency (RF) depletion spectroscopy is employed to characterize the internal state distribution of the molecules. Fourth, extensive Monte-Carlo trajectory simulations of the electrostatic guiding allow for the properties of the molecules directly after the buffer-gas cell to be inferred from the signal at the end of the guide.\\
With these tools, we are able to characterize the buffer-gas cooling in various aspects. We demonstrate addressing of all significantly populated internal states of the guided molecules emerging from our cryogenic buffer-gas source, operated in different regimes, boosted~\cite{Maxwell2005,MotschBoosting} or supersonic~\cite{Hutzler2011}. We applied this method to different species of polyatomic symmetric-top polar molecules, fluoromethane, $\rm{CH_3F}$, and 3,3,3-trifluoropropyne, $\rm{CF_3CCH}$, and made comparison of their cooling processes. We provide clear evidence of the efficient control over the internal cooling of the guided molecular beams through varying the buffer-gas-cell temperature and the buffer-gas density, demonstrating the possibility of cooling far below the cell temperature in the supersonic regime. The detailed study of the buffer-gas cooling reveals two interesting phenomena. First, comparing the rotational with the translational temperature, we provide evidence that rotational cooling takes place more efficiently than translational cooling for the CH$_3$F-He system in the low He density regime. Second, the measurements provide indications of dependence of the collisional relaxation rate on the rotational states of the molecules.\\
Our paper is structured as follows. In Section~\ref{Section_Method} we present a brief review of symmetric-top molecules, and then describe the principle of our method for rotational-state detection of guided molecules. Section~\ref{Section_ExpSetup} describes the experimental set-up. The characterization and analysis of the radio-frequency depletion measurements are described in Section~\ref{Section_ExpResults}. The Monte Carlo trajectory simulations used to calculate the guiding efficiencies for all relevant rotational states are described in Section~\ref{Section_MonteCarlo}. Finally, in Section~\ref{Section_Discussion} we present the results from the comprehensive characterization of our buffer-gas source. There we analyze the cell's output for different operating regimes and for different molecular species, and draw inferences on the cooling mechanisms in the buffer-gas cell.
\section{Method for rotational-state detection}
\label{Section_Method}
One of the key features in this work is the implementation of a new method for state detection, which is a pivotal tool for the characterization of cold molecular beams from cryogenic buffer-gas sources. Several techniques for state detection of cold molecules have been used so far~\cite{vanVeldhoven2002,MotschDepletion,Bertsche2010,Barry2011,Patterson2012,Twyman2014,Gloeckner2015a}. Among them, depletion methods, as adopted in previous experiments carried out by our group~\cite{MotschDepletion,Gloeckner2015a}, are particularly advantageous since they are applicable to a large range of molecules and avoid the difficulties of the light-induced fluorescence (LIF) and resonance-enhanced multi-photon ionization (REMPI) detection techniques. Along this line, we have extended the depletion technique by implementing rotational-state detection adapted to the cold guided molecular beams from a cryogenic buffer-gas source, using resonance RF depletion spectroscopy in a parallel-plate capacitor.\\
\subsection{The symmetric-top molecule}
\label{Subsection_SymmTopMol}
To understand our method for rotational-state detection, first we briefly review the relevant properties of symmetric-top molecules, which have been used in the current study. Their rotational states can be fully described by three quantum numbers~\cite{Wollrab1967}, the total angular momentum, $J$, its projection on the molecule's symmetry axis, $K$, ($K=-J, ..., J$), and its projection on a laboratory-fixed axis, $M$, ($M=-J, ..., J$). Hereinafter the rotational states will be designated as $|JKM\rangle$ or $|JK\rangle$, depending on the need to specify the quantum number $M$.\\
In the absence of external electric fields, the energy of a symmetric-top molecule in the rigid-rotor approximation is expressed as $E_{J,K}=h[BJ(J+1)+(A-B)K^2]$, where $h$ is Planck's constant, and $A$ and $B$ are the rotational constants of a symmetric-top molecule. The presence of an external electric field splits every $|JK\rangle$ state (Stark effect) into $2J+1$ $M$-sublevels, corresponding to all possible projections of $\rm{\mathbf{J}}$ on the electric-field axis. The first-order Stark splitting is given by the expression $E^{(1)}=-\mu\mathcal{E}\frac{KM}{J(J+1)}$, where $\mu$ and $\mathcal{E}$ stand for the permanent electric dipole moment of the molecule and the electric-field strength, respectively. Sublevels with a positive Stark shift are referred to as low-field-seeking states (states that are guided). Since states $|J\,-KM\rangle$ are degenerate with states $|JK\,-M\rangle$ under inversion symmetry, we ignore hereinafter the sign of $K$ and adopt the convention that states with positive $M$ are low-field-seeking. An example of the Stark effect is presented in Figure~\ref{Fig_ExpSetup}(a) for the $|1,1\rangle$ and $|2,1\rangle$ states of $\rm{CH_3F}$. For $|JK\rangle$ states with $K=0$, the first-order Stark splitting is zero. In this case, low-field-seeking states due to the second-order (quadratic) Stark shift exist, if they satisfy the condition $J(J+1)>3M^2$, which derives from the expression for the quadratic Stark shift $E^{(2)}=-\frac{\mu^2\mathcal{E}^2}{2hB}\left[\frac{3M^2-J(J+1)}{J(J+1)(2J-1)(2J+3)}\right]$ for $K=0$~\cite{Wollrab1967}. The expression of $E^{(2)}$ for the case $M=0$ is identical to the above one, except the quantum number $M$ in the expression is replaced by $K$. Thus states with $M=0$ can also be low-field-seeking states.\\
\subsection{Radio frequency depletion spectroscopy}
\label{Subsecton_RFDSpectroscopy}
Our method for rotational-state detection is based on state-selective addressing and elimination of molecules from a guided population by applying an RF field resonant to the DC Stark splitting in a homogeneous field. This leads to a depletion of the measured signal and reveals the relative state population.\\
Molecules emerging from the cryogenic source populate different rotational states $|JK\rangle$. Only molecules in low-field-seeking $M$-sublevels, however, are confined in the guide. Subsequently, the guide is interrupted by a parallel-plate capacitor creating a homogeneous offset electric field. At low field, this causes equidistant Stark splitting of the $M$-sublevels for a given $|JK\rangle, K\neq0$ state. The magnitude of the splitting is unique for most of the low-lying $|JK\rangle$ states for a given offset field (see Figure~\ref{Fig_ExpSetup}(a)). This allows for $|JK\rangle$ states to be addressed individually. An RF field which is resonant to the Stark splitting between the $M$-sublevels of the rotational state $|JK\rangle$ we want to address is applied to the capacitor, and transfers the molecules into different $M$-sublevels. Those molecules which land in non-guidable states are eliminated from the guiding after the capacitor, and this leads to a depletion signal on the detector.\\
To achieve a good control over the resulting $M$-substate distribution, we broaden the RF signal with white noise and hence eliminate coherent effects in the population transfer process. The molecules are thereby equally redistributed among all the $M$-sublevels within the given $|JK\rangle$ rotational state (Figure~\ref{Fig_ExpSetup}(a)), when the transitions are power-saturated. To a first approximation, the depletion ratio in this case is given by the fraction of molecules converted from guidable to non-guidable states, and this is the ratio between the number of non-guidable $M$-sublevels and the total number of $M$-sublevels, i.e., $\frac{J+1}{2J+1}$ for a linear Stark shift. The fraction of guided molecules populating the probed rotational state is therefore obtained by dividing the magnitude of the saturated depletion by this ratio. In addition, a decrease in signal results also from transferring molecules from a guidable state with a large Stark shift to one with a small Stark shift, as the latter are more prone to losses in the guide. Thus the simplified depletion ratio needs to be corrected including the subtle effects of guiding efficiency of molecules in different $M$-sublevels, which are described in detail in Section~\ref{Section_MonteCarlo}. 
\section{Experimental Set-up}
\label{Section_ExpSetup}
\begin{figure}[t]
\centering
\includegraphics[width=1.0\linewidth]{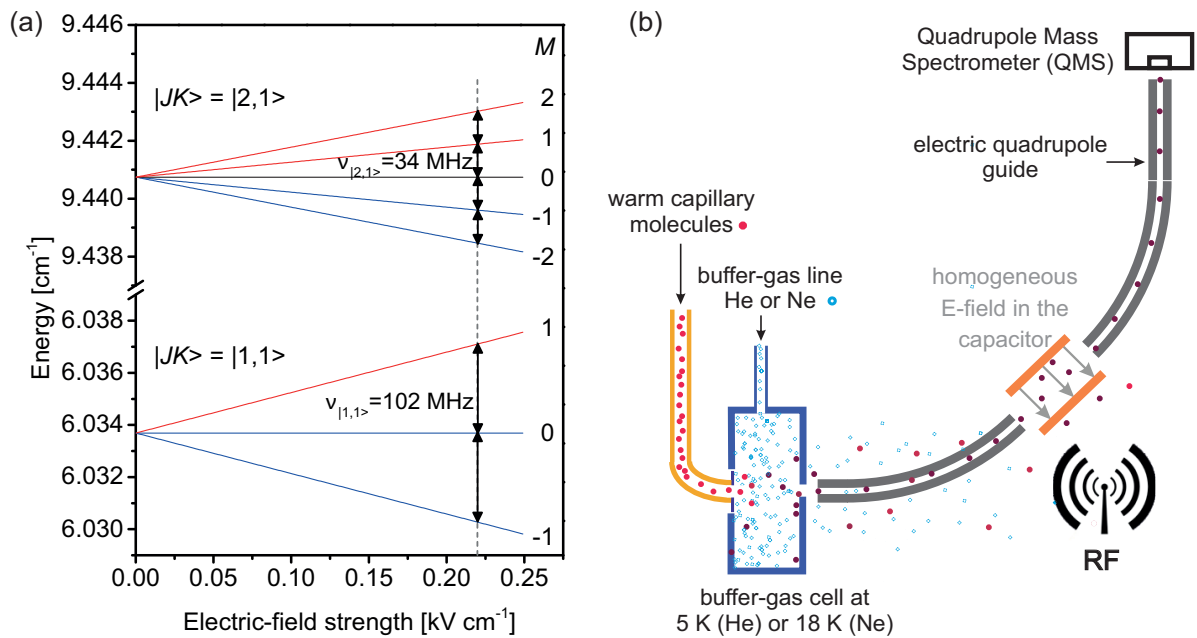}
\caption{(a) Stark splittings as a function of the applied electric field for two rotational states of $\rm{CH_3F}$, $|1,1\rangle$ and $|2,1\rangle$: red lines show guidable sublevels, and blue lines show non-guidable ones. The $|2,1,0\rangle$ state is weakly guidable due to its second-order Stark shift. The vertical dashed line designates the applied homogeneous electric field ($220\,\rm{V\,cm^{-1}}$) used in the radio-frequency scans shown in Figure~\ref{Fig_RFSpectroscopy}(a). The arrows show the magnitude of the Stark splitting corresponding to the applied electric field. This splitting is unique for most low-lying $|JK\rangle$ states. (b) Scheme of the experimental set-up.}
\label{Fig_ExpSetup}
\end{figure}

The scheme of the experimental set-up is shown in Figure~\ref{Fig_ExpSetup}(b). In our experiment, buffer-gas cooling takes place in a cryogenic cell where a continuous flow of polyatomic polar molecules is mixed with a continuous flow of helium or neon at cryogenic temperatures of $5\,\rm{K}$ or $18\,\rm{K}$, respectively. The exact cell temperature can be monitored by diode sensors and controlled by an electric heating block. The gas inflow is monitored by pressure gauges and regulated by needle valves along the corresponding gas lines. After the cold molecules leave the cell, they are captured by a quadrupole guide, which delivers them to a detector or to further experiments. The experimental set-up is an upgraded version of the set-up described previously~\cite{Sommer2009} to enable RF depletion spectroscopy on the guided molecules.

Our buffer-gas set-up can be operated in two different regimes, boosted and supersonic. The supersonic regime is achieved for buffer-gas densities typically one to two orders of magnitude higher compared to those used in the boosted regime, and is characterized by a large number of collisions at the cell output orifice and within a short distance downstream from it. The high buffer-gas densities result in a high pressure gradient across the nozzle, and consequently, in an adiabatic expansion, leading to further cooling of the gas emerging from the cell~\cite{Hutzler2011}. The boosted regime is an intermediate between the supersonic and the effusive (i.e., collisionless) regime, where a few collisions occur near the nozzle, and thereby the slowest molecules are eliminated from the beam. The two regimes require different geometries. For the boosted regime, the guide is distanced about $2\,\rm{mm}$ from the cell nozzle. In the case of the supersonic regime, that spacing is {\it ca.} $20\,\rm{mm}$ to allow enough distance for the supersonic beam formation. The cell exit aperture used in our experiments has a diameter of $2\,\rm{mm}$. The maximum electric field that can be achieved between the guiding electrodes is about $100\,\rm{kV}\,\rm{cm^{-1}}$. We use a quadrupole mass spectrometer (QMS) to detect the molecules after they leave the guide.\\
The capacitor for driving RF transitions is built of two parallel ($2\,\rm{cm}\times2\,\rm{cm}$) plates with a $2.75\,\rm{mm}$ spacing between them. A bias voltage from a few tens of volts up to a few thousand volts is applied to the capacitor, providing a homogeneous electrostatic field between the plates. The capacitor is separated from the two guide segments by gaps of $1\,\rm{mm}$. The plates are made from circuit boards. One of the plates is a plain board, whereas the other one features a T-shaped microstrip used to apply the RF field. As the flat end of the T-shaped microstrip is aligned with the molecules' flow axis, the RF field is expected to be predominantly perpendicular to the homogeneous field in the molecules' transit region. This configuration enables driving of $\Delta M=\pm1$ transitions.\\
\section{Radio frequency depletion measurements for state detection}
\label{Section_ExpResults}
\begin{figure}[t]
\centering
\includegraphics[width=1.0\linewidth]{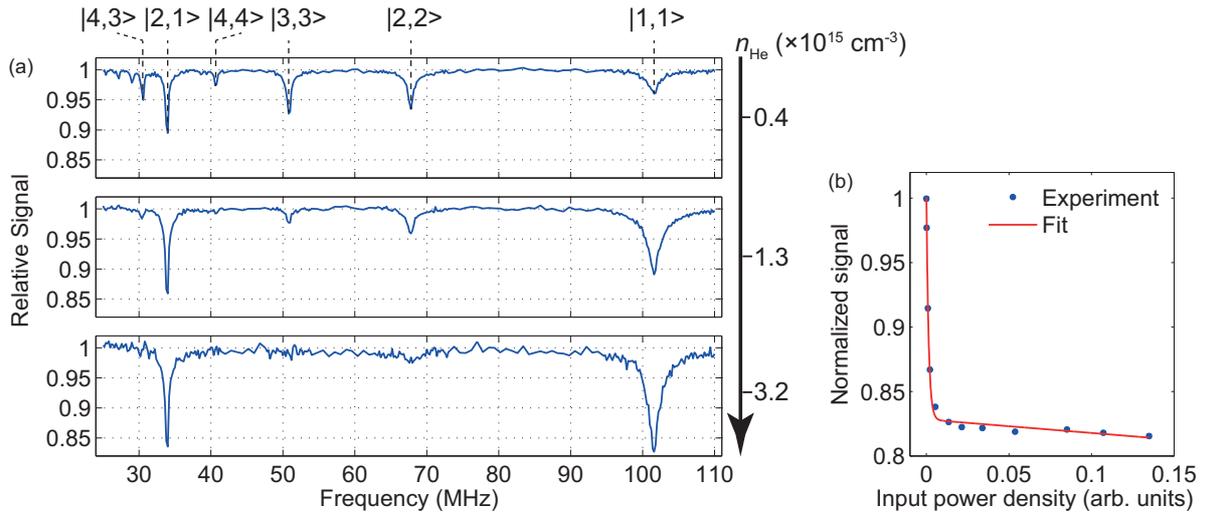}
\caption{(a) RF depletion spectra of $\rm{CH_3F}$ for different He densities in the buffer-gas cell ($T_{\rm{cell}}=6.7\,\rm{K}$). The most prominent dips are assigned to the corresponding $|JK\rangle$ rotational states. (b) Depletion saturation measurement for the $|1,1\rangle$ state of $\rm{CH_3F}$. The fit to the experimental points is a sum of an exponential and a linear power dependence (For details, see text).}
\label{Fig_RFSpectroscopy}
\end{figure}

In this Section we present the characterization and the analysis of radio-frequency depletion measurements of buffer-gas-cooled and electrically guided molecules, and the derivation of their rotational-state distribution based on these measurements. Figure~\ref{Fig_RFSpectroscopy}(a) shows the relative signal of guided helium-buffer-gas-cooled fluoromethane, $\rm{CH_3F}$, as a function of the applied RF frequency, for different buffer-gas densities in the cell, $3.7\times10^{14}\rm{cm^{-3}}$, $1.3\times10^{15}\rm{cm^{-3}}$, and $3.2\times10^{15}\rm{cm^{-3}}$, and for a cell temperature of $6.7\,\rm{K}$. The applied homogeneous electric field of $220\,\rm{Vcm^{-1}}$ was the same for all three scans. The linear Stark splittings for different $|JK\rangle$ rotational states of fluoromethane have been calculated for the applied homogeneous electric field, and, on this basis, the observed features in the depletion spectrum have been assigned. The assignments of the most prominent dips are shown in the figure. The broadening of the lines is attributed to the inhomogeneity of the electric field, which leads to a linewidth proportional to the scanning frequency.\\
Figure~\ref{Fig_RFSpectroscopy}(a) shows a clear change in the dip pattern and in the relative dip depth as the buffer-gas density is changed. Increasing the buffer gas density in the cell increases the number of collisions and therefore leads to a lower rotational temperature. This effect is clearly visible, as the depletion signal corresponding to the low-lying rotational states $|1,1\rangle$ and $|2,1\rangle$ increases, while the signal from the other states of higher rotational energy decreases, and eventually vanishes.\\
The magnitude of the depletion signal depends not only on the thermal population of the probed state, but also on the applied RF power. To eliminate this effect, we performed RF power scans for each $|JK\rangle$ rotational state to obtain its saturated depletion level. Coherent effects, which lead to Rabi oscillations between the $M$-sublevels and therefore hinder an equal redistribution of population among all $M$-sublevels, were eliminated by broadening the RF signal to $5\,\rm{MHz}$ with white noise (the white noise is not applied in the frequency scan in Figure~\ref{Fig_RFSpectroscopy}(a)). 
An example of an RF power scan is shown in Figure~\ref{Fig_RFSpectroscopy}(b) for the $|1,1\rangle$ state of $\rm{CH_3F}$, where the relative signal is plotted as a function of the applied RF power spectral density. The data were fitted with a sum of an exponential and a linear function. The exponential term is attributed to resonant driving of the transition of interest, while the linear decay term is attributed to off-resonant driving of transitions in other states taking place in the non-homogeneous-field regions near the quadrupole guide, as well as to transitions driven by higher harmonics produced by the RF amplifier that are not perfectly filtered out. The sought value of the saturated depletion for each of the considered rotational states is taken to be the amplitude of the exponential function. For $\rm{CH_3F}$ we have performed saturation scans only for states with $J\leq4$ under the reasonable assumption that states with higher $J$ rotational number have a negligible thermal population below $10\,\rm{K}$.\\
Results from the RF depletion measurements of $\rm{CH_3F}$ for $T_{\rm{cell}}=6.4\,\rm{K}$ and $n_{\rm{He}}=1.6\times10^{15}\,\rm{cm^{-3}}$ are shown in Table~\ref{Table_Populations}. The values listed in the second column show the measured depletion signal for a given rotational state $|JK\rangle$ from the RF depletion scans. The nominal depletion ratio, $\frac{J+1}{2J+1}$ (discussed in Section~\ref{Subsecton_RFDSpectroscopy}), is listed in the third column. The more accurate simulated depletion ratio obtained from the trajectory simulations described in detail in the next Section is listed in the fourth column of Table~\ref{Table_Populations}. The relative population of each of the rotational states $|JK\rangle$ present in the guided beam can be deduced from the ratio of the measured depletion signal and the simulated depletion ratio. Note that the sum of the state populations is $\sim 100\,\%$, indicating our ability to account for the entire population of rotational states of $\rm{CH_3F}$ present in the guided beam.\\

\begin{table}
\centering
\newcolumntype{d}[1]{D{.}{\cdot}{#1} }
\begin{tabular}{ld{2}lld{1}}
\toprule \toprule
\multicolumn{1}{c}{State}&\multicolumn{1}{c}{Depletion}&\multicolumn{1}{c}{Nominal DR}&\multicolumn{1}{c}{Simulated DR}&\multicolumn{1}{c}{Population}\\
\multicolumn{1}{c}{$|JK\rangle$}&\multicolumn{1}{c}{(\%)}&&\multicolumn{1}{c}{(\%)}&\multicolumn{1}{c}{(\%)}\\
\midrule
$|1,1\rangle$&$16.96\,(0.14)$&$2/3$& $66.7$ &$25.4\,(0.2)$\\
$|1,1\rangle$&$16.96\,(0.14)$&$2/3$& $66.7$ &$25.4\,(0.2)$\\
$|2,1\rangle\,[|3,2\rangle$]&$19.07\,(0.11)$&$3/5\,[4/7]$& $57.3\,[58.8]$ &$32.8\,(0.2)$\\
$|3,1\rangle$&$5.08\,(0.12)$&$4/7$& $54.3$ &$9.4\,(0.2)$\\
$|4,1\rangle$&$1.02\,(0.13)$&$5/9$& $55.6$ &$1.8\,(0.2)$\\
$|2,2\rangle$&$1.88\,(0.11)$&$3/5$& $62.2$ &$3.0\,(0.2)$\\
$|1,0\rangle$&$12.32\,(0.15)$&$2/3$& $66.7$ &$18.5\,(0.2)$\\
$|2,0\rangle$&$5.6 \,(0.4)$&$2/3$& $66.0$ &$8.5\,(0.6)$\\
\midrule
Total&$62.6 \,(0.5)$& & &$\bf{99.4\,(0.8)}$\\
\bottomrule \bottomrule
\end{tabular}
\caption{Experimental relative populations of the $|JK\rangle$ rotational states of $\rm{CH_3F}$ derived from saturated depletion signals measured for each of the rotational states at $T_{\rm{cell}}=6.4\,\rm{K}$ and $n_{\rm{He}}=1.6\times10^{15}\,\rm{cm^{-3}}$. Nominal and simulated depletion ratios (DR), are defined and explained in the text. The values in parentheses of the depletion and the population columns are the statistical errors. The $|3,2\rangle$ state in the square brackets in the second row shares the same Stark splitting with the $|2,1\rangle$ state, and its contribution to the depletion signal is also taken into account. The corresponding depletion ratios for the $|3,2\rangle$ state are also given in square brackets in the third and fourth column.}
\label{Table_Populations}
\end{table}

\section{Monte Carlo trajectory simulations and guiding efficiencies}
\label{Section_MonteCarlo}
To characterize the buffer-gas source and to quantify the effect of buffer-gas cooling for different cell operating regimes, it is necessary to determine the rotational-state and velocity distributions of the molecules upon emerging from the cryogenic buffer-gas cell. To retrieve this information from the measurements at the end of the guide, we perform comprehensive Monte Carlo trajectory simulations, which yield the guiding efficiencies for the molecules in the quadrupole guide. In this Section, we describe the basic principle and some crucial elements of the simulations.\\
\begin{figure}[t]
\centering
\includegraphics[width=1.0\linewidth]{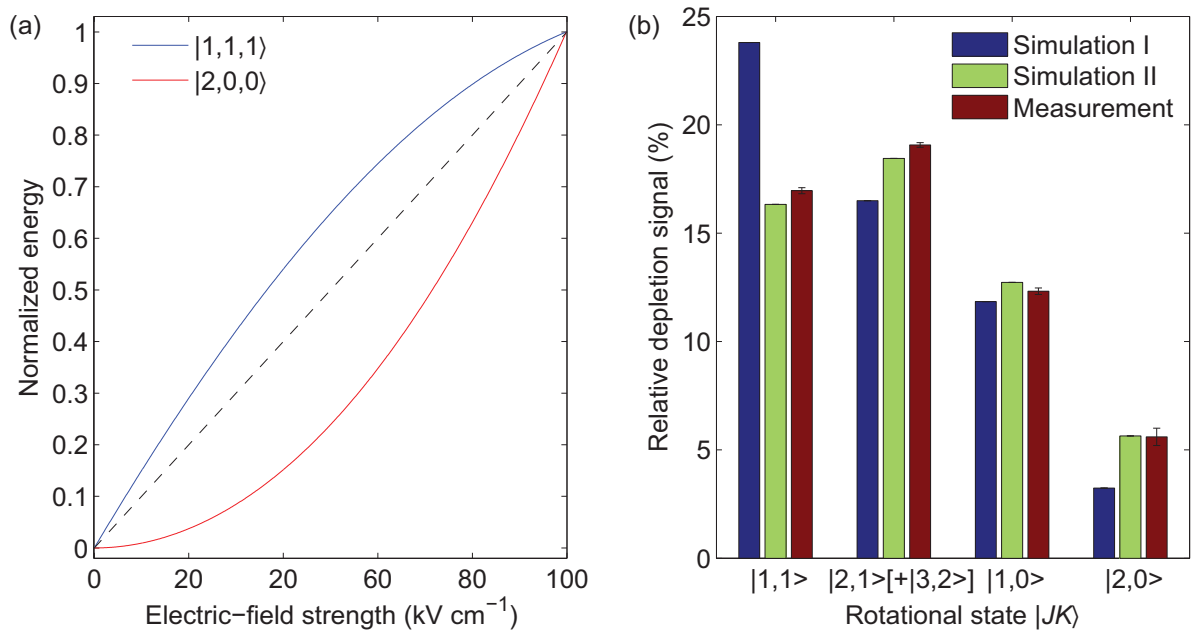}
\caption{(a) Normalized exact Stark-shift curves as a function of the electric field for states with different types of Stark-shift curvatures: convex ($|1,1,1\rangle$ rotational state of $\rm{CH_3F}$) and concave ($|2,0,0\rangle$ rotational state of $\rm{CH_3F}$). They were obtained based on diagonalizing the rigid-rotor Hamiltonian. The linear approximation to the Stark-shift curve is shown for comparison (black dashed line). (b) Demonstration of the necessity of using the exact Stark-shift curves: comparison between the relative depletion signals for simulation I , simulation II, and the measurement. The simulated depletion signals are taken at the fitted $T_{\rm{rot}}$, with the fitting procedure explained in details in Section~\ref{Subsection_TrotTcell}. The major difference between simulations I and II is that the former uses only linear approximations to the Stark curves while the latter uses the exact Stark curves. Very good agreement with the experimental results is obtained only for simulation II.}
\label{Fig_StarkCurves}
\end{figure}
The electric-field distribution along the entire molecule guide is calculated using the SIMION software package, and, on this basis, the molecular trajectories from the cell output to the QMS are simulated. In particular, this includes the two 20~cm-radius 45-degree bends, the capacitor plates, and the divergence of the molecular beam between the end of the guide and the QMS, as shown in Figure~\ref{Fig_ExpSetup}(b). To determine the longitudinal velocity distributions, experiments and simulations are also performed for the case of a continuous guide, in which the capacitor plates are replaced by a straight piece of guide, as here the capacitor plates are not necessary. The effect of applying an RF field to the capacitor plates is simulated by evenly redistributing the population of a given $|JK\rangle$ state across all of its $M$-sublevels in the middle of the capacitor. The ratio of the simulated signals with and without reshuffling the population across the M-sublevels gives the simulated depletion ratio.\\
In the following, we describe the initial conditions and the method for the trajectory simulations. The transverse spatial distribution of the molecules at the guide input is taken to be flat, and the transverse energy distribution is taken as the Boltzmann distribution ($\sim T_{\rm{cell}}$) truncated by the transverse acceptance of the guide ($\sim1K$ trap depth). The initial longitudinal velocity, $v_{\parallel}$, is varied, allowing a guiding efficiency as a function of $v_{\parallel}$ to be extracted from the simulation. Dividing the measured $v_{\parallel}$-distribution after the guide by the corresponding guiding efficiencies yields the initial $v_{\parallel}$-distribution before the guide. Assuming this initial $v_{\parallel}$-distribution is the same for all internal states, we simulate the guiding efficiency for every relevant $|JKM\rangle$ state with the above-determined initial $v_{\parallel}$-distribution taken as an input. Once these state-dependent guiding efficiencies become available, we could recover the initial rotational-state distribution before the guide (see Figure~\ref{Fig_LongVeloDistrib}(a)) from the depletion measurements described in Section~\ref{Section_ExpResults}. One subtlety here is that the simulations for the $v_{\parallel}$ distribution and for the rotational-state distribution need each other's results as their inputs. This interdependence can be resolved by realizing that the $v_{\parallel}$-dependent guiding efficiency is only weakly influenced by the state composition, and the two simulations can be performed iteratively. In practice, we start with a state distribution corresponding to the Boltzmann distribution at $T_{\rm{cell}}$ in order to determine the initial $v_{\parallel}$-distribution at the guide input. This in turn helps us calculate the theoretical depletion values for various states between the capacitor plates, assuming a Boltzmann distribution in the cell. By fitting the measured RF depletion data to the so-obtained theoretical values, we can determine the rotational temperature of the molecules exiting the cell. This last assumption of reaching a Boltzmann distribution in the cell at a fitted temperature is validated by the results shown in Section~\ref{Subsection_TrotTcell}.\\
To achieve a good agreement between the simulated and the experimental results, a key requirement is to take into account the exact dependence of the Stark shift on the electric field, rather than using a linear or quadratic approximation. As shown in Figure~\ref{Fig_StarkCurves}(a), the Stark shift for certain states deviates strongly from a linear approximation for fields up to 100~kV/cm, as present in the experiment. Employing the exact Stark-shift curve dramatically improves the match between the measured and simulated depletion signal, as shown in Figure~\ref{Fig_StarkCurves}(b). In particular, including the exact Stark curve decreases the simulated depletion signal for states with a convex Stark shift and increases the simulated depletion signal for states with a concave one, with a less pronounced effect for states with a Stark shift more closely following a linear behaviour, as might be expected.\\
\section{Analysis and discussion}
\label{Section_Discussion}
Sections~\ref{Section_ExpResults} and \ref{Section_MonteCarlo} have described the measurement and the simulation methods. In this Section we apply these tools for a comprehensive characterization of the buffer-gas-cooling process in the cryogenic cell. First, we explore in Sections~\ref{Subsection_TrotTcell} and \ref{Subsection_TrotBGdensity} the influence of cell parameters on the buffer-gas cooling. Next, we investigate in Sections~\ref{Subsection_Trifluropropyne} and \ref{Subsection_Supersonic} the cooling of different molecular species and in the supersonic operating regime of the cell. At the end, in Sections~\ref{Subsection_TrotTtrans} and \ref{Subsection_RotStateDependence} we discuss and draw conclusions on the cooling effect for different degrees of freedom and for different rotational states.\\
\subsection{Dependence of the rotational temperature on the cell temperature}
\label{Subsection_TrotTcell}

\begin{figure}[t]
\centering
\includegraphics[width=1.0\linewidth]{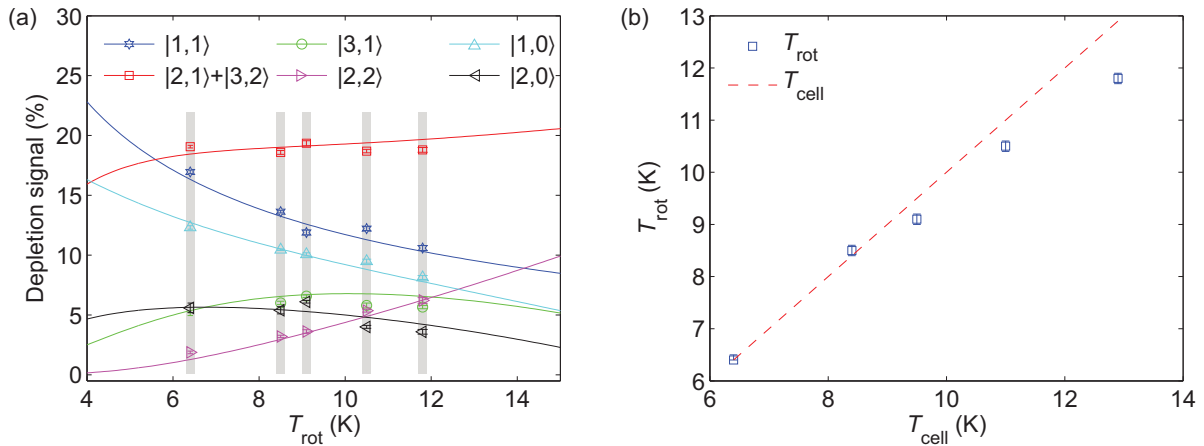}
\caption{(a) Measured depletion signals (data points) for different rotational states $|JK\rangle$ of $\rm{CH_3F}$ and different buffer-gas cell temperatures at a fixed He flux of $\Phi_{\rm{He}}=2\times10^{17}\,\rm{s^{-1}}$ and simulated depletion signals (lines) for different rotational temperatures for different rotational states. The grey columns group sets of measurements performed at the same $T_{\rm{cell}}$. Measurements at five different cell temperatures have been performed. The rotational temperature of the molecules leaving the buffer-gas cell was determined by columnwise simultaneous fitting of the measured depletion signals to the calculated curves. (b) Relation between the measured cell temperature, $T_{\rm{cell}}$, and the fitted rotational temperature, $T_{\rm{rot}}$, for $\rm{CH_3F}$.}
\label{Fig_TrotFit}
\end{figure}

The first aspect we consider is the dependence of the rotational temperature on the buffer-gas cell temperature. Here, we would like to address two questions: first, whether the rotational states of molecules obey a Boltzmann distribution in the cell; second, how $T_{\rm{rot}}$ compares to $T_{\rm{cell}}$. Figure~\ref{Fig_TrotFit}(a) shows the measured depletion (in percentage of the total signal) for different rotational states $|JK\rangle$ of $\rm{CH_3F}$ for different cell temperatures, $T_{\rm{cell}}$, along with their theoretical predictions (solid curves). The measurements were done at a fixed He flux of $\Phi_{\rm{He}}=2\times10^{17}\,\rm{s^{-1}}$ (corresponding to a density of $n_{\rm{He}}=1.6\times10^{15}\,\rm{cm^{-3}}$ at $T_{\rm{cell}}=6.4\,\rm{K}$ for our cell geometry). The data for each $T_{\rm{cell}}$ are grouped in a separate vertical column in Figure~\ref{Fig_TrotFit}(a). The theory curves are based on the assumption of a Boltzmann distribution for the rotational states at the cell (as mentioned in Section~\ref{Section_MonteCarlo}), with $T_{\rm{rot}}$ being an independent variable on the horizontal axis. The guiding efficiencies are simulated using the method described in Section~\ref{Section_MonteCarlo}. In particular, we have included all 64 guidable $|JKM\rangle$ states of $\rm{CH_3F}$ (rotational constants $A=155\,\rm{GHz}$ and $B=26\,\rm{GHz}$), with rotational energies below $72\,\rm{K}$ ($50\,\rm{cm^{-1}}$). For a source at $10\,\rm{K}$ those states account for $99.7\,\%$ of the thermal population. Note that $T_{\rm{rot}}$ is the only fit parameter in Figure~\ref{Fig_TrotFit}(a), which sets the horizontal position of each column via a least-$\chi^2$ fit to the theory curves. The good agreement between measurement and theory confirms the assumption that a Boltzmann distribution is reached for the rotational states in the cell, at the He flux applied in this particular measurement. Moreover, the plot in Figure~\ref{Fig_TrotFit}(b) also shows a relatively good agreement between $T_{\rm{rot}}$ and $T_{\rm{cell}}$. This, however, is not a universal effect, but rather a consequence of the applied He flux. In fact, as will be shown in the next Section, by varying the He density in the cell, we also vary $T_{\rm{rot}}$.
\subsection{Dependence of the rotational temperature on the buffer-gas density}
\label{Subsection_TrotBGdensity}
\begin{figure}[t]
\centering
\includegraphics[width=1.0\linewidth]{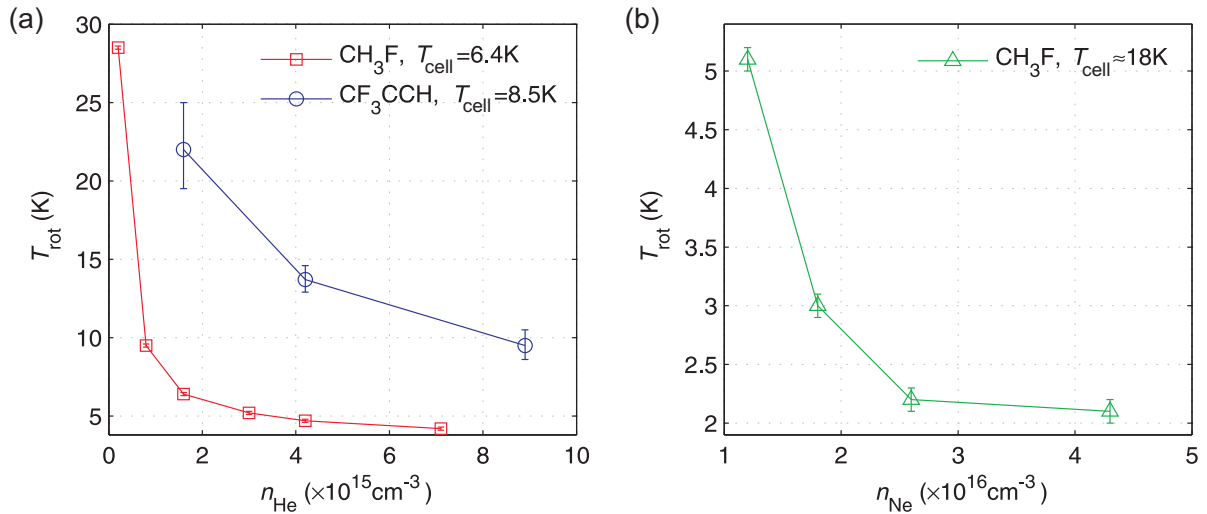}
\caption{(a) Rotational temperatures of two molecular species, $\rm{CH_3F}$ and $\rm{CF_3CCH}$, as a function of the He buffer-gas density in a cell operated in the boosted regime. The cell temperatures were kept constant, $6.4\,\rm{K}$ and $8.5\,\rm{K}$, for $\rm{CH_3F}$ and $\rm{CF_3CCH}$, respectively. (b) Rotational temperature of $\rm{CH_3F}$ as a function of the Ne buffer gas in a cell operated in the supersonic regime. The cell temperature was kept fixed at $18\,\rm{K}$.}
\label{Fig_HeDensity}
\end{figure}

As already pointed out qualitatively (see Section~\ref{Section_ExpResults}), increasing the buffer-gas density in the cell increases the cooling capacity, and, consequently, results in a better cooling of the molecules. To quantify this effect, we determined the rotational temperature of $\rm{CH_3F}$ for different He densities ranging from $2.2\times10^{14}\rm{cm^{-3}}$ to $7.1\times10^{15}\rm{cm^{-3}}$ at a fixed cell temperature of $T_{\rm{cell}}=6.4\,\rm{K}$. The dependence is plotted in Figure~\ref{Fig_HeDensity}(a) (red squares). We observe cooling of rotational states to near the cell temperature for minimal He density below $1\times10^{15}\rm{cm^{-3}}$ for our cell geometry. It is also particularly interesting to note that for buffer-gas densities above $2\times{10^{15}\rm{cm^{-3}}}$, the rotational temperature of the $\rm{CH_3F}$ molecules leaving the cell is lower than the cell temperature. This evidences further cooling as a result of the adiabatic expansion of the gas upon streaming out of the cell~\cite{Hutzler2011,Barry2011}. The higher the buffer-gas density, the stronger the effect of the adiabatic cooling. For a He density of $7.1\times10^{15}\rm{cm^{-3}}$, the rotational temperature is determined to be $(4.2\pm0.1)\,\rm{K}$.\\

\subsection{Cooling of a heavier molecule}
\label{Subsection_Trifluropropyne}
In addition to varying the cell parameters, we also investigated the effect of the molecular species on the cooling process. Here we cool the rotational degrees of freedom of a heavier molecule, 3,3,3-trifluoropropyne, $\rm{CF_3CCH}$ (mass $m=94\,\rm{u}$, rotational constants $A=5.7\,\rm{GHz}$ and $B=2.8\,\rm{GHz}$), and compare the results with those for $\rm{CH_3F}$ (mass $m=34\,\rm{u}$). The at least 10-fold decrease in energy spacing ($\propto A,\,B$) for $\rm{CF_3CCH}$ compared to that of $\rm{CH_3F}$ not only greatly reduces the signal per state in the measurement, but also requires more energy levels to be included in the simulation. In this case we have performed trajectory simulations including the exact Stark curves for all 2869 guidable $|JKM\rangle$ levels below 72~K (corresponding to 99.7\% thermal population at 10~K). We have derived the rotational temperature of $\rm{CF_3CCH}$ for different He densities in a similar range as for $\rm{CH_3F}$, from $1.6\times10^{15}\rm{cm^{-3}}$ to $8.9\times10^{15}\rm{cm^{-3}}$, and for a fixed buffer-gas cell temperature of $T_{\rm{cell}}=8.5\,\rm{K}$, as shown in Figure~\ref{Fig_HeDensity}(a) (blue circles). The higher cell temperature in this case results from the maintained higher temperature of the molecule feed line, as $\rm{CF_3CCH}$ has a considerably higher freezing point than $\rm{CH_3F}$. The trend of better cooling for higher buffer-gas densities is clearly observed. Nevertheless, even for the largest He density applied here, the rotational temperature remains slightly above the cell temperature, i.e., the molecules are not fully thermalized. Various effects play a role in determining the thermalization rate. A possible reason for the worse cooling of the rotational degrees of freedom for $\rm{CF_3CCH}$ compared to $\rm{CH_3F}$ for similar buffer-gas densities includes the larger density of rotational states populated by $\rm{CF_3CCH}$ requiring more collisions to thermalize.\\
\subsection{Rotational temperatures in the supersonic regime}
\label{Subsection_Supersonic}
So far we have only discussed the buffer-gas cooling in the boosted regime. We have also measured the rotational temperature of the $\rm{CH_3F}$  molecules leaving the buffer-gas cell operated in the supersonic regime. The supersonic beams are formed when a large number of collisions take place near the cell nozzle during expansion into a vacuum. During this process, the particles convert their internal energy into a kinetic one whereby they get colder~\cite{Levy1980}. In this case, we used Ne as a buffer gas since at $T_{\rm{cell}}=18\,\rm{K}$ it has a supersonic flow speed below $200\,\rm{m\,s^{-1}}$, lower compared to the supersonic flow velocity of He at $6\,\rm{K}$, which makes guiding more feasible. The other benefit of using Ne is the better pumping of the background gas. The dependence of the measured rotational temperature, $T_{\rm{rot}}$, as a function of the buffer-gas density is shown in Figure~\ref{Fig_HeDensity}(b). Here we demonstrate that the supersonic expansion leads to a reduction of the rotational temperature by almost an order of magnitude below the cell temperature, allowing us to cool molecules down to $(2.1\pm0.1)\,\rm{K}$.\\

\begin{figure}[t]
\centering
\includegraphics[width=1\linewidth]{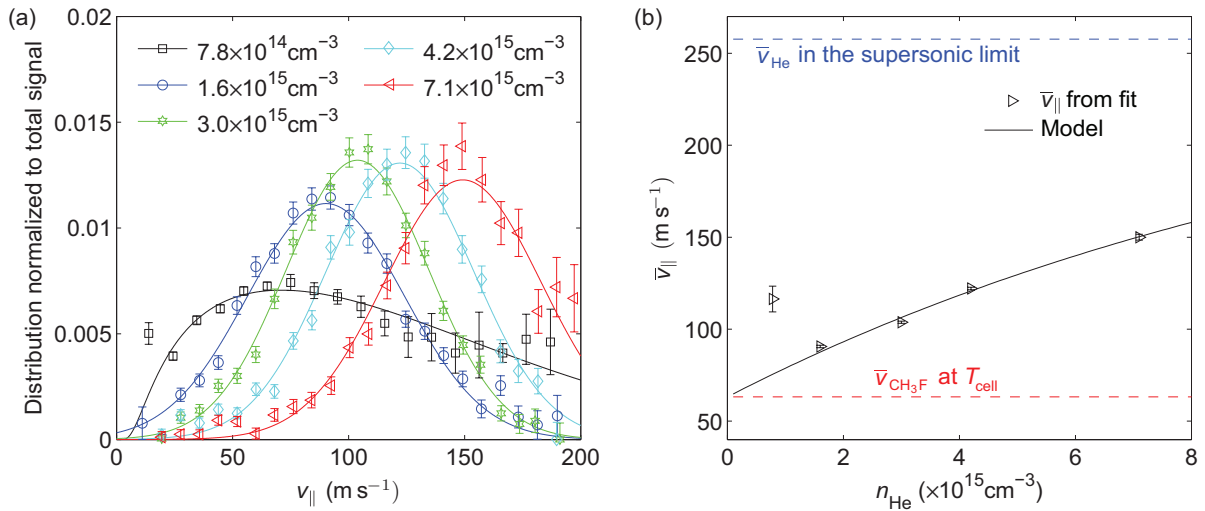}
\caption{(a) Derived longitudinal velocity distribution for $\rm{CH_3F}$ at the input of the guide for $T_{\rm{cell}}=6.4\,\rm{K}$ and different He densities, along with their fits. (b) Mean longitudinal velocity, $\bar{v}_{\parallel}$, from the fits as a function of the corresponding $n_{\rm{He}}$. The solid curve is based on a momentum transfer model described in the text.}
\label{Fig_LongVeloDistrib}
\end{figure}

\subsection{Comparison between rotational and translational temperatures}
\label{Subsection_TrotTtrans}
In addition to characterizing the rotational-state distribution and measuring $T_{\rm{rot}}$ of the molecules emerging from the cell, it is also important to retrieve their longitudinal velocity distribution ($v_{\parallel}$-distribution), determine the translational temperature, and compare it to the rotational one. The procedure of deducing the $v_{\parallel}$-distributions at the cell output is explained in Section~\ref{Section_MonteCarlo}. Figure~\ref{Fig_LongVeloDistrib}(a) shows the $v_{\parallel}$-distributions of $\rm{CH_3F}$ for the same cell parameters as those used in the $T_{\rm{rot}}$ measurement in Section~\ref{Subsection_TrotBGdensity}. The solid curves in Figure~\ref{Fig_LongVeloDistrib}(a) are fits to the distributions. Gaussian fits centred at peak velocities have been applied to all but the lowest-density distribution ($n_{\rm{He}}=7.8\times10^{14}\,\rm{cm}^{-3}$). The latter has been fitted with a Maxwell distribution corrected for the boosting effect~\cite{MotschBoosting}. The use of a different fit function is substantiated by the quantitative model below.\\
To further analyze the changes in the $v_{\parallel}$-distribution, we plot in Figure~\ref{Fig_LongVeloDistrib}(b) the mean longitudinal velocity, $\bar{v}_{\parallel}$, as a function of $n_{\rm{He}}$. The value of $\bar{v}_{\parallel}$ is extracted from the fits in Figure~\ref{Fig_LongVeloDistrib}(a). Its increase at higher $n_{\rm{He}}$ is understood to result from collisions between molecules and He atoms in the vicinity of the nozzle~\cite{Maxwell2005}. The lighter He atoms move at a higher speed and kick the heavier molecules from behind whereby the molecules get accelerated in the forward direction. To quantify this effect, we have built the following model based on the momentum transfer for elastic collisions. From conservation of energy and momentum, the change in the molecules' velocity per head-on collision is $\Delta v_{\parallel}=\beta(v_{\rm{He}}-v_{\parallel})$, where $\beta=\frac{2m_{\rm{He}}}{m_{\rm{mol}}+m_{\rm{He}}}$ with $m_{\rm{mol(He)}}$ being the mass of the molecule (He atom), and $v_{\rm{He}}$ is the longitudinal velocity of the He atom. As the He density in the nozzle region is typically much higher than the molecule density, we assume $v_{\rm{He}}$ to be constant and approaching the supersonic velocity, which sets the upper bound for the mean velocity of He in the limit of high density~\cite{Levy1980}. This leads to $v_{\parallel}(l)=\bar{v}_{\rm{He}}-(\bar{v}_{\rm{He}}-v_{0})exp(-\beta l)$, where $l$ is the number of head-on collisions in the molecular-beam direction, $\bar{v}_{\rm{He}}$ is the centre velocity of a He supersonic beam at $T_{\rm{cell}}$, and $v_{0}$ is the initial longitudinal velocity of the molecules. $l$ is related to $n_{\rm{He}}$ by $l=dg\sigma n_{\rm{He}}$, where $d=2$mm is the nozzle diameter setting a rough estimate of the distance where collisions take place, $\sigma$ is the $\rm{CH_3F}$-He elastic collision cross-section, and $g$ is a proportionality factor on the order of unity taking into account the fact that only collisions in the direction normal to the nozzle plane contribute to the change in $v_{\parallel}$. Assuming the molecules have translationally thermalized to $T_{\rm{cell}}$ before arriving at the nozzle, and hence $v_{0}$ is approximately their thermal mean velocity at $T_{\rm{cell}}$, our model (the solid curve in Figure~\ref{Fig_LongVeloDistrib}(b)) describes very well the dependence of $\bar{v}_{\parallel}$ with respect to $n_{\rm{He}}$ with the exception of the lowest-density point. The only fit parameter in the model is $g\sigma=2.0\times10^{-15}\,\rm{cm}^2$, which agrees within one order-of-magnitude with the typical $\rm{CH_3F}$-He collision cross-section at this temperature range~\cite{Willey1988}. Moreover, the fact that the lowest-density data stays above the model curve shows that the translational degree of freedom of molecules is not yet thermalized to $T_{\rm{cell}}$ at this point, in line with the finding in the $T_{\rm{rot}}$ measurement at low He densities (see Section~\ref{Subsection_TrotBGdensity}). This also justifies using a different fit function for the lowest-density $v_{\parallel}$-distribution (Figure~\ref{Fig_LongVeloDistrib}(a)).\\
In the following we discuss the translational temperature of molecules, $T_{\rm{tr}}$, and compare it to the rotational one. While the width of a Gaussian fit to the $v_{\parallel}$-distribution defines $T_{\rm{tr}}$ in the high-density limit, where molecular beams approach the supersonic regime, it provides only a lower bound to $T_{\rm{tr}}$ in the low-density limit where a Maxwellian distribution describes $v_{\parallel}$ better, as shown in Figure~\ref{Fig_LongVeloDistrib}(a). Thus we assign a lower bound to $T_{\rm{tr}}$ of $15.7\pm1.7\,\rm{K}$ for the distribution at $n_{\rm{He}}=7.8\times10^{14}\,\rm{cm}^{-3}$ based on its width, while its mean kinetic energy corresponds to $28\pm3\,\rm{K}$. In comparison, $T_{\rm{rot}}$ at this density is found to be $9.5\pm0.1\,\rm{K}$ (the second point in Figure~\ref{Fig_HeDensity}), which is clearly lower than the value of $T_{\rm{tr}}$. Having a colder $T_{\rm{rot}}$ than $T_{\rm{tr}}$ might not be surprising for the given system because of the large mass imbalance between $\rm{CH_3F}$ and He, which makes cooling of the centre of mass motion inefficient. On the other hand, the $v_{\parallel}$-distribution for the highest density in Figure~\ref{Fig_LongVeloDistrib}(a) ($n_{\rm{He}}=7.1\times10^{15}\,\rm{cm}^{-3}$) has a width corresponding to $4.7\pm0.4\,\rm{K}$, which is very similar to $T_{\rm{rot}}=4.2\pm0.1\,\rm{K}$ found in Section~\ref{Subsection_TrotBGdensity}. Presumably, in the high-density limit, $T_{\rm{tr}}$ and $T_{\rm{rot}}$ should converge due to the sufficient number of collisions.\\
\subsection{Rotational-state-dependent thermalization rates}
\label{Subsection_RotStateDependence}

\begin{figure}[t]
\centering
\includegraphics[width=1.0\linewidth]{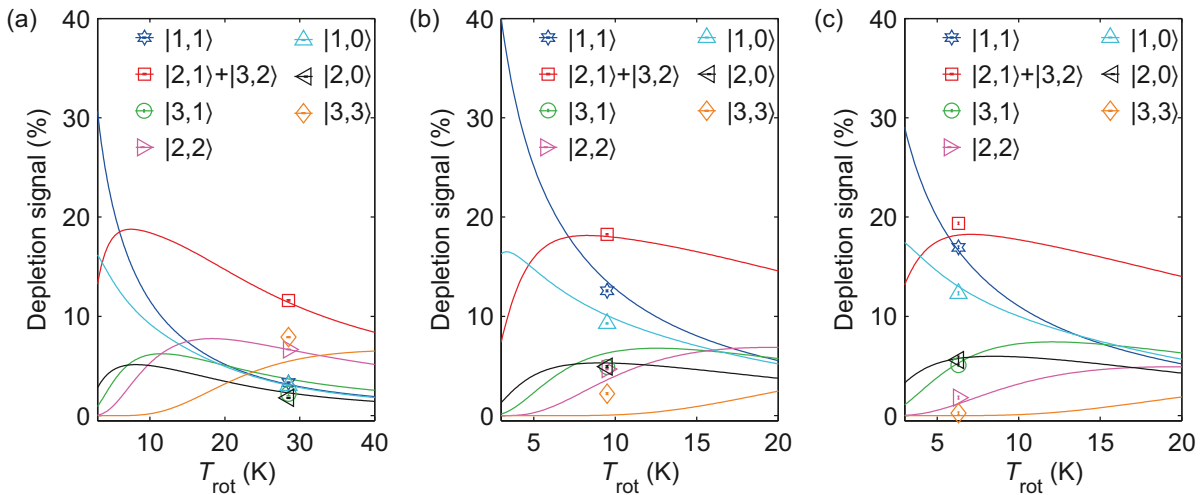}
\caption{Measured depletion signals (data points) for different rotational states of $\rm{CH_3F}$ and different He densities, (a) $n_{\rm{He}}=2.2\times10^{14}\rm{cm^{-3}}$, (b) $n_{\rm{He}}=7.8\times10^{14}\rm{cm^{-3}}$, and (c) $n_{\rm{He}}=1.6\times10^{15}\rm{cm^{-3}}$, at a fixed cell temperature $T_{\rm{cell}}=6.4\,\rm{K}$, and simulated depletion signals for different rotational states and different rotational temperatures (lines) for each of the applied He densities.}
\label{Fig_JKThermalization}
\end{figure}

In Section~\ref{Subsection_TrotBGdensity} we have investigated the dependence of fitted $T_{\rm{rot}}$ on the He density in the cell. In this Section we take a closer look at the state distribution in the low He density limit. Figure~\ref{Fig_JKThermalization} shows a comparison between the measured depletion signals and the corresponding theory curves in increasing order of He density, $n_{\rm{He}}=2.2\times10^{14}\rm{cm^{-3}}$, $7.8\times10^{14}\rm{cm^{-3}}$, and $1.6\times10^{15}\rm{cm^{-3}}$ and at $T_{\rm{cell}}=6.4\,\rm{K}$, corresponding to the first three points of the $\rm{CH_3F}$ data in Figure~\ref{Fig_HeDensity}(a). Similarly to Figure~\ref{Fig_TrotFit}(a), the data are grouped in columns, and the only fit parameter is the horizontal position of the column, which determines $T_{\rm{rot}}$. The results clearly demonstrate that with increasing He density, the rotational states are better thermalized, and the fitted $T_{\rm{rot}}$ approaches $T_{\rm{cell}}$. Moreover, at lower He density, the larger deviation between the measured depletion signals and the theoretical curves also indicates that the internal states are not fully thermalized. In particular, we would like to point out that the depletion signals for the $|33\rangle$ state (orange diamonds in Figure~\ref{Fig_JKThermalization}) appear far above their theoretical curves, and would correspond to a much higher rotational temperature if the $|33\rangle$ state were fitted alone. This is an indication that the $|33\rangle$ state thermalizes more slowly than the other states. Based on nuclear spin statistics, states with $K$ being a multiple of 3 (ortho-$\rm{CH_3F}$) are not interconvertible with states of $K$ not being a multiple of 3 (para-$\rm{CH_3F}$)~\cite{Wollrab1967} via buffer-gas collisions. Hence, $\rm{CH_3F}$ from the $|33\rangle$ state can only relax to states with $K=0$. Since the rotational energy scales as $K^2$, a transition of the type $K=0\leftarrow K=3$ requires 3 times more energy to be taken away than a transition of the type $K=1\leftarrow K=2$ does. Thus we attribute the above-observed effect to the larger energy difference between the initial and the final state, which makes the $|33\rangle$ state thermalize at a lower rate than the other probed states.\\ 
\section{Summary and outlook}
We have developed a versatile and robust method for rotational and translational thermometry of cold molecules emerging from a cryogenic buffer-gas source. The method consists of three independent tools, time-of-flight measurements yielding the longitudinal velocity distribution, resonant radio-frequency rotational-state-selective detection of guided molecules, capable of addressing $100\,\%$ of their population, and a complete Monte-Carlo simulation package resolving the guiding efficiencies for all present states.\\
The obtained results make possible the full characterization of buffer-gas sources and provide insight into the cooling processes in a cryogenic cell operated in different regimes. Studying those mechanisms, we have established an efficient control over the rotational and translational cooling of guided molecular beams by tuning the buffer-gas-cell temperature and the buffer-gas density. We have demonstrated rotational cooling below the cell temperature for the strongly boosted and supersonic regimes, achieving rotational temperatures more than an order of magnitude lower than the cell temperature. Interesting physical phenomena have also been observed. Comparing the rotational with the translational temperature, we have shown evidence of faster thermalization of the rotational degrees of freedom for the $\rm{CH_3F}$-He collisions at the limit of low $n_{\rm{He}}$. We have also observed manifestations of rotational-state-dependent cooling rates, and we have provided an explanation of this effect.\\
The presented thermometry method is a very general one, which can be applied to a vast variety of molecules. It is particularly useful for probing cold molecules that are not amenable to other techniques, e.g., LIF or REMPI. Moreover, the thorough understanding of the cooling mechanisms in a cryogenic buffer-gas cell, provided by the new method, is the key to optimizing the production of internally and translationally cold molecules.\\

\textbf{Acknowledgement} We would like to thank Ferdinand Jarisch for the technical support in the lab.

\section{References}

\bibliographystyle{unsrt}

\end{document}